# Cryogenic microwave frequency combs based on quantum paraelectric superconducting resonators

Prasad Muragesh[1 #], Harikrishnan Sundaresan [1 #], and Madhu Thalakulam [1 *]

[1] School of Physics, Indian Institute of Science Education & Research Thiruvananthapuram, Thiruvananthapuram Kerala 695551 India

[#] equally contributed to the work
[*] madhu@iisertvm.ac.in

## Abstract

A frequency comb, known for its precision as an "optical ruler", features an evenly spaced spectral pattern. While these combs are vital in photonic quantum technologies, their microwave counterparts are now highly sought after for cryogenic quantum technologies, including semiconducting and superconducting qubits and quantum electrical metrology, which mainly operate in the microwave regime. However, microwave combs are still largely underexplored, and typically rely on complex, high-power optical systems incompatible with the low-power, cryogenic on-chip quantum technologies. In this manuscript, we present an all-electrical, on-chip, cryogenic microwave frequency comb on Strontium Titanate ($SrTiO_3$), exploiting its Pockels-like effect in its quantum paraelectric phase. Our device, utilizing a superconducting microwave cavity, generating the frequency comb via cavity phase modulation enabled by the field-induced effective $\chi(2)$ of $SrTiO_3$. The ability to continuously vary the dielectric constant of $SrTiO_3$ by the application of electric field, in its quantum paraelectric phase, makes it possible to control the comb's operating frequency range. The exceptionally high dielectric constant of $SrTiO_3$, > 20,000 in its quantum paraelectric state, enables an ultra-miniature design and on-chip integration with cryogenic quantum technologies.

## I. Introduction.

A frequency comb is a spectral structure comprising numerous discrete, equally spaced frequency lines, visually analogous to the teeth of a comb [1]. Traditionally, mode-locked lasers have served as the primary means of generating these devices [2,3]. Four-wave mixing in micro-resonators [4] and electro-optic modulation [5] are also being explored for optical frequency comb production. Optical frequency combs play a critical role in photonic quantum technologies, including astronomical spectrograph calibration [6], high-precision metrology [7], and optical atomic clocks [8,9]. Recently, their microwave counterparts have attracted increasing attention, motivated by advancements in quantum technologies within the microwave domain [10]. Microwave frequency combs enable applications such as high-precision radar ranging [11], and enhanced wireless communication [12]. There is a distinct demand for low-power microwave frequency combs, especially in the field of cryogenic quantum technology [13,14]. In semiconducting and superconducting quantum computing platforms, qubit manipulation and readout depend upon precise, low-power microwave signals [15]. Additionally, quantum voltage and current standards also relies on highly accurate microwave signals [16–18]. In contrast, the development of an on-chip, all-electrical, low-power cryogenic microwave frequency comb remains largely unrealised.

Microwave frequency combs reported largely exploits the interaction of lasers with non-linear media, such as Kerr non-linearity [19–21], $\chi^{(2)}$ non-linearity [22,23], and four-wave mixing [24–26]. Recent advancements in cavity opto-mechanics have enabled the creation of optomechanical [27–30] and electromechanical [31–33] frequency combs where the cavity field is mechanically or electrically modulated to generate comb lines. Electro-optic frequency combs utilizing the $\chi^{(2)}$ non-linearity offer greater tuneability compared to cavity optomechanical combs. The reliance on extensive optical components and higher power budgets makes most of these approaches incompatible with cryogenic quantum applications and transitioning entirely to the microwave circuit-based design is the path forward.

Reports on cryogenic frequency comb generation predominantly exploits superconducting microwave cavities coupled with non-linear elements, such as Josephson junctions [34–36], SQUID [14,37], or nanomechanical oscillators [31–33]. Additionally, superconducting resonators composed of non-linear kinetic inductance materials [38] have been explored for comb generation. However, these systems generally have limited tuneability in terms of free spectral range (FSR). Furthermore, the physical dimensions of these combs, dictated by the wavelength of microwaves, are not fully optimized for integration with cryogenic on-chip quantum devices (see Supplemental Materials S-1 for a short survey on microwave frequency combs).

A promising approach to achieving an all-electric, cryogenic microwave frequency comb is to embed a microwave cavity within an inherently non-linear medium that features an electrically tuneable response function at cryogenic temperatures. In this regard, Strontium Titanate ($SrTiO_3$ or STO), a quadratic electro-optic material [39–41] with a large third-order non-linear susceptibility, $\chi^{(3)}$, at cryogenic temperatures makes an attractive choice. Though STO is predicted to be in the ferroelectric phase for temperatures < 40 K, the transition is suppressed by zero-point fluctuations of the soft transverse optical phonon modes and remains in an unpolarized state, the so-called quantum paraelectric state: a highly polarizable, incipient ferroelectric state stabilized by quantum fluctuations [42]. In this state, biasing STO crystal with an electric field, can induce an effective $\chi^{(2)}$ behavior $\chi_{eff}^{(2)} \approx \chi_0^{(2)} + 3E_b\,\chi_0^{(3)}$ [43] with a very high Pockels coefficient [44]. In this work, we exploit the effective $\chi^{(2)}$ non-linearity in the quantum paraelectric phase at sub-Kelvin temperatures for the generation of an on-chip, all-electrical microwave frequency comb on STO. The device utilizes a high-$Q$ coplanar waveguide (CPW) microwave cavity fabricated on an STO crystal with (001) orientation. The cavity is excited at its resonance, while a modulation signal applied across the centre conductor and ground-plane alters the dielectric constant, producing phase modulation that generates the frequency comb. Our design ensures complete tunability of the resonant frequency, modulation frequency, and FSR, resulting in a fully tuneable all-electrical frequency comb. Exploiting strategies such as flip-chip technology or by depositing STO directly onto the target substrate, our strategy presented here brings up the prospect of on-chip integration with circuit-based quantum device architectures[44–47].

## II. Results & Discussions

### A. Induced $\chi^{(2)}$ non-linearity in STO in the quantum paraelectric phase

Fig. 1(a) shows a schematic diagram of the experimental setup; refer Supplemental Materials S-2 for details. An aluminium CPW cavity realized on STO is excited at its resonant frequency $f_c$ with a continuous-wave microwave signal. Simultaneously, a low-frequency electrical modulation signal of frequency $f_m$ is applied across the dielectric gap between the centre conductor and the ground plane through a bias tee. The reflected signal is amplified further and measured using a spectrum analyser. The modulating field at $f_m$ induces a time-varying change in the permittivity of STO and therefore on the effective dielectric constant $\varepsilon_{eff}$ of the cavity. The resulting time varying change in the phase velocity $v_p = {}^{c}/_{\sqrt{\epsilon_{eff}}}$ induces phase modulation of the signal whose amplitude takes the form

$$V(t) = A_c \cdot \cos(2\pi f_c t + \beta \cdot \sin(2\pi f_m t)) \quad (1)$$

which represents a frequency comb; a set of equally spaced spectral lines centred around $f_c$ in the frequency-domain with an FSR of $f_m$. $A_c$ and $\beta$ are the carrier amplitude and phase modulation index [49].

Fig. 1(b) shows an optical image of the aluminium superconducting CPW cavity on $500\ \mu m$ single crystal STO. The CPW on STO, designed to have an impedance of $\sim 1\ \Omega$, forms a microwave cavity due to the impedance mismatch with $50\ \Omega$ RF circuitry. An external bias-tee enables the application of both DC and AC voltages between the centre conductor and ground plane. Details on the device architecture and fabrication can be found in the Supplemental Materials S-3. The cavity, $1.5\ mm$ in lithographic length, is designed to have its fundamental and first-resonant modes at 1 $GHz$ and 2 $GHz$, respectively below a temperature of 4 K. This extreme miniaturization is made possible due to the very high dielectric constant of STO, $> 20{,}000$ in its quantum paraelectric state [50]. The simulated electric field distribution, obtained via eigenmode analysis (Ansys HFSS), for the first-resonant mode is shown in the bottom-panel of Fig.1(b), while that of the fundamental mode is provided in Supplemental Materials S-4. All experiments in this manuscript are conducted with the first-harmonic mode.

The narrow gap between the centre conductor and the ground plane on the cavity translates a small voltage into an intense electric field across the dielectric gap. For example, 1 V generates an electric field $\sim 100\ kV/m$. Fig. 1(c) shows the single-port reflectivity $S_{11}$, at 500 mK for the first harmonic mode, $\sim 2.04\ GHz$, with no DC or AC voltages applied; the fundamental mode is at 955 $MHz$ (not shown). We estimate a cavity Q-factor of $\sim 2.8 \times 10^5$ from Fig. 1(c).

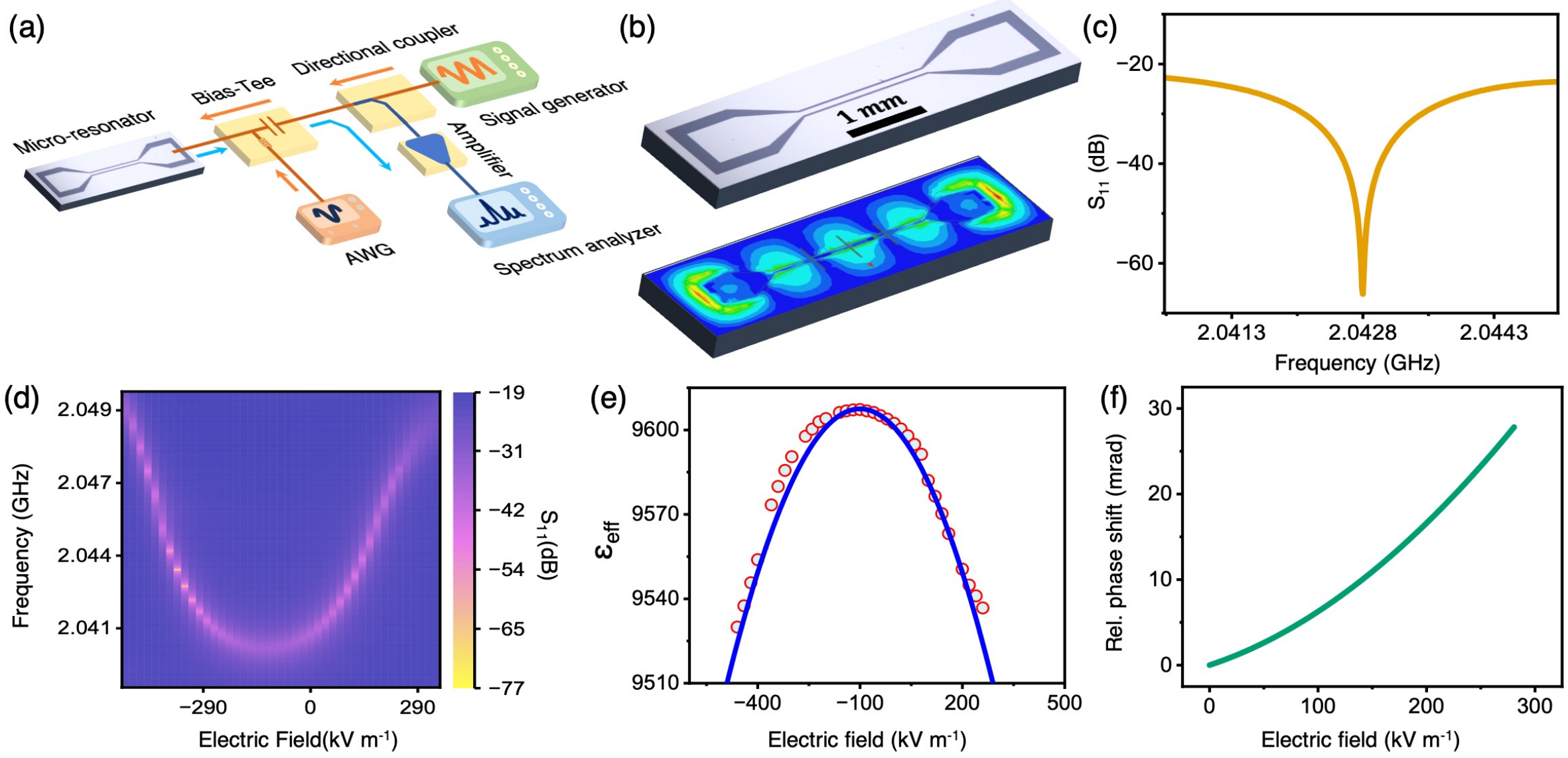

**Fig.1. Phase modulation in an STO superconducting microwave cavity. (a)** Schematic diagram of the measurement setup. **(b)** Upper panel: Optical micrograph of the superconducting cavity designed for 1 $GHz$ fundamental mode. Lower panel: simulated electric field distribution of the first resonance mode ($\sim 2\ GHz$), obtained via eigenmode analysis using Ansys HFSS. **(c)** Measured reflection coefficient $|S_{11}|$ for the first resonant mode of the cavity, corresponding to the simulated mode in (**b**). **(d)** Plot of $|S_{11}|$ against the DC electric field showing variation in the resonant frequency under the DC bias. **(e)** Change in the effective dielectric constant extracted from **(d)** as a function of the applied electric field (red open circles), blue line corresponds to the fit to LGD model (equation 2). **(f)** Computed relative phase shift ($\delta$) as a function of the applied field, extracted from **(e).**

Fig. 1 (d) shows a plot of the $|S_{11}|$ in the vicinity of the resonance as a function of DC electric field across the dielectric gap. The asymmetric variation of the resonant frequency about the zero-field suggest the presence of remnant polarization in the crystal [51]. Under the application of the electric field, the resonant frequency shifts from the zero-field value due to the field induced modification of the dielectric constant [39,40,50]. We find that the resonator's bandwidth ($Q$-factor) increases (decreases) with applied DC voltage. The quasi-Debye mechanisms [52] and the field-induced polarization introduce anharmonicity into the energy landscape leading to nonlinear dielectric behaviour, dielectric relaxation and consequently an increase in the loss-tangent at higher DC voltages [53]. An application of 5 $V$, corresponding to a field of $\sim 500\ kV/m$, shifts the resonant frequency by $\sim 10\ MHz$. The effective dielectric constant of STO is extracted using a comparative approach [50]; the resonant frequency of an identical resonator realized on a Sapphire substrate is compared to that on the STO [Supplemental Materials S-5].

Fig. 1(e) shows the dielectric constant versus electric field, extracted from the resonant frequencies shown in Fig. 1(d). The shift in the maximum dielectric constant in Fig. 2 (e) towards non-zero fields is believed due to remnant polarization. Oxygen defects, commonly found in STO, can create local dipole moments, in-turn creating micro-polar regions, while the bulk is still quantum paraelectric state [54–56]. In addition, STO also suffers from the possible dead-layer formation with non-zero polarization, whose non-uniformity also contributes to the overall remanent polarization [57–59]. In this regard, the effect of remnant polarization on the high-frequency performance of a varactor has been reported [60].

The dielectric constant versus electric field for STO is given by the Landau-Ginzburg-Devonshire (LGD) model [61],

$$\epsilon_{eff}^{STO}(E) = 1 + \frac{\epsilon_{eff}^{STO}(0)}{[1+(E-E_c/E_0)^2]^{1/3}} \tag{2}$$

Where, $\epsilon_{eff}^{STO}(E)$ is the effective dielectric constant at electric field $E$, $E_c$ is the coercive field and $E_0$ is a parameter related to the dielectric tunability. The blue line in Fig. 1(e) shows a fit to the model from which we can extract the change in dielectric constant against a given applied voltage. The relative phase shift (δ) against dielectric constant variation can be expressed as

$$\delta = \frac{2\pi}{\lambda}\,\Delta\left(\sqrt{\varepsilon_{eff}}\right) \times L \tag{3}$$

where $\lambda$ is the free space wavelength of the microwave signal, $L$ is the cavity length and $\Delta\left(\sqrt{\varepsilon_{eff}}\right)$ represents the variation in the dielectric constant against applied electric field [Supplemental Materials S-6]. Fig. 1(f) shows the calculated δ versus applied electric field. We find that application of a 5 $V$ DC bias across the centre conductor and the ground plane induces a $\delta \sim 70\ mrad$. Presence of the high-Q cavity in our setup, further enhances the degree of phase modulation by trapping the microwave signal over many cycles. Therefore, the $\beta$ gets enhanced by a factor of $\sqrt{Q}$, $\beta \approx \delta \cdot \sqrt{Q}$, making the system an efficient phase modulator even for low applied electric fields. This is an important aspect especially for cryogenic quantum circuits where the power dissipation is a crucial factor affecting the device operation. In our experiments we confine to modulation signal of amplitudes ranging from a few $kV/m$ to $62.5\ kV/m$, which corresponds to a maximum effective dielectric constant change of ∼ 8 relative to the zero-bias one. From Fig. 1(e), this corresponds to a $\beta \sim 1.05$ rad.

## B. Cavity-enhanced microwave comb generation

Fig. 2 shows the microwave frequency comb generated in our setup described in Fig. 1(a) and 1(b). We employ a two-tone excitation scheme: a carrier tone at the resonant frequency produced by an RF signal generator and a modulation tone provided by an arbitrary waveform generator (AWG). These tones interact via the effective second-order nonlinearity ($\chi^{(2)}$) stemming from remnant polarization in the STO crystal (as shown in Fig. 1(d)). The modulation frequency we supply along with the resonant drive frequency modulates the dielectric constant of the STO. Resulting parametric oscillations phase-modulate the resonant frequency with the pump frequency, resulting in frequency mixing and forming a comb—sidebands spaced by the modulation frequency around the carrier. Fig. 2(a) illustrates the spectrum of the generated frequency comb observed in the cavity's reflected power. For this demonstration, the carrier tone is set to $\sim 2\ GHz$, while a $10\ kHz$, $\sim 0.625\ V_{pp}$ sinusoidal modulation signal is applied across the centre conductor and ground plane. Throughout this study, the input carrier wave power is maintained at $\sim -24\ dBm$ unless specified otherwise. The resulting FSR of $10\ kHz$ matches the modulation frequency precisely, producing a comb with 16 clearly discernible sidebands symmetrically distributed around the carrier frequency. The amplitude distribution of these sidebands mirrors the profile of $S_{11}$ near the resonant frequency, underscoring the intrinsic connection between the system's resonance characteristics and the structure of the generated frequency comb. This implies that the comb can be centred around any desired frequency by appropriately designing the cavity geometry. Fig. 2(b) shows the simulated frequency comb using equation (1) with a modulation frequency of $10\ kHz$ and $\beta = 1.05\ rad$, showing an excellent agreement with the experimental data in Fig. 2(a).

Next, we inspect the spectral purity of the comb utilising phase noise measurements, shown in Fig. 2(c). We find that the phase noise remains well below −100 dBc/Hz for offset frequencies above 1 kHz for the carrier wave, indicating that the device contribution to the phase noise is almost negligible. For the first comb line, the phase noise also drops below −75 dBc/Hz for all offset frequencies above 2 Hz. Both of these metrics are comparable to the performance benchmarks reported for microwave frequency combs [62,63]. A summary of phase noise measurements are provided in the Supplemental Materials S-7. We have measured the equidistant spacing of the comb lines and the drift of the spectrum over a duration of 30 hours and find that the spectrum is stable and no measurable drift is observed (Supplemental

Materials S-8 and S-9). The low phase noise, long-term stability and equidistant comb-lines is indicative of a phase-coherent frequency-comb.

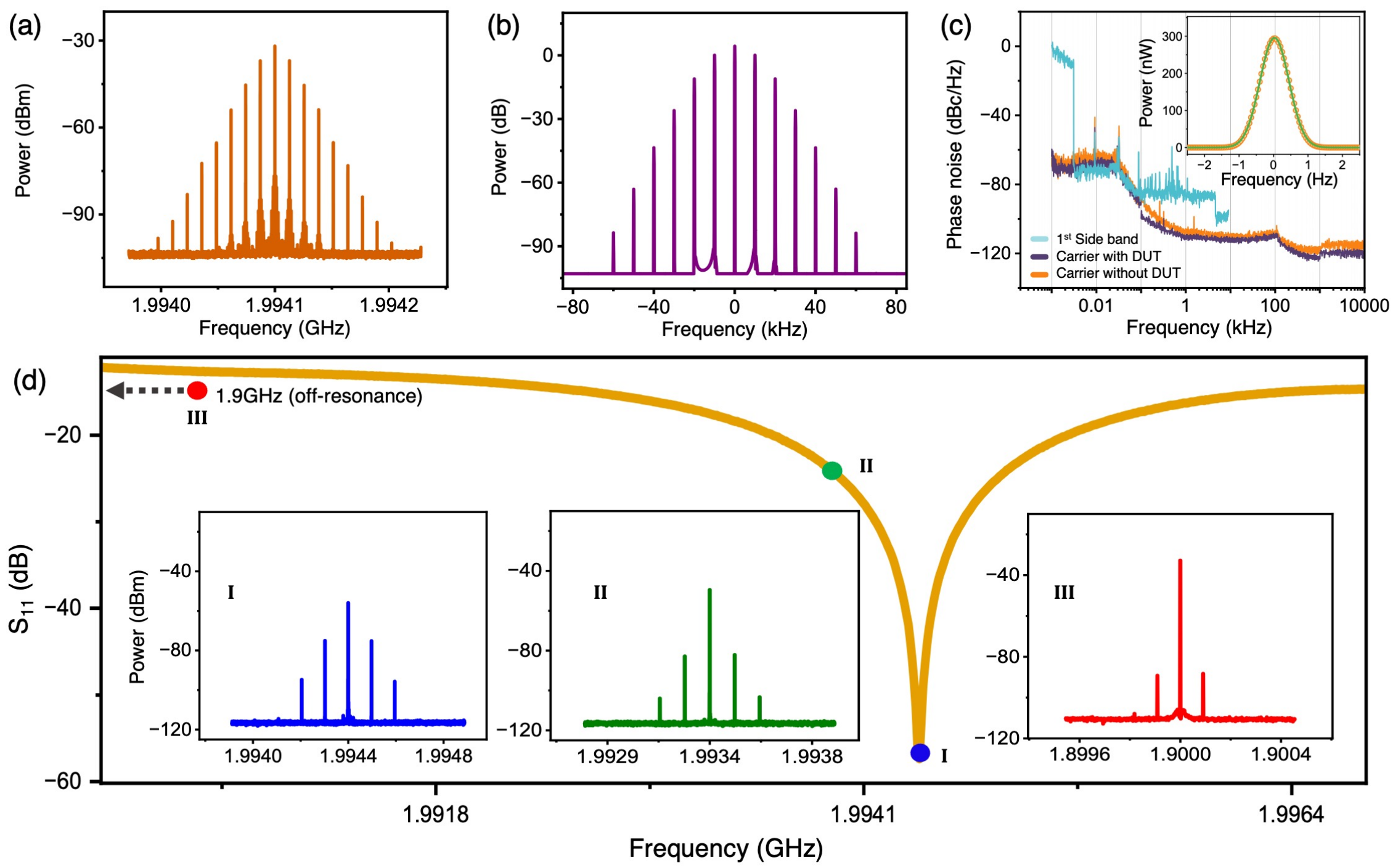

**Fig.2. Frequency comb generation using two-tone excitation. (a),** Experimentally measured frequency comb generated from the STO device using a sinusoidal modulation at 10 $kHz$ around a carrier frequency of ~2 $GHz$. **(b)** Simulated frequency comb based on phase modulation using Equation (1), with parameters matched to the experimental conditions in **(a).** Apparent broadening of the base of a few comb lines arises from numerical artifacts associated with the finite-duration time trace used in the simulation and the subsequent Fast Fourier Transform (FFT). **(c)** Phase noise measurements for the carrier wave with (violet) and without (orange)the device under test (DUT) and for the 1st sideband (teal)). The inset shows a magnified view of the first sideband from panel **(a)**, with a spectral resolution ~0.996 $Hz$ obtained from the Voigt fit. The green curve represents the Voigt fit, while the orange circles correspond to the experimental data. **(d),** Frequency combs generated by detuning the carrier wave around the resonance frequency (Points **I**, **II** and **III**) with a 10 $kHz$, 0.125 $V_{pp}$ modulation. Insets **I**, **II**, and **III** show the corresponding combs at each detuning point.

The inset of Fig. 2(c) provides a magnified view of the first sideband relative to the carrier wave in Fig. 2(a) and find that the spectral resolution is primarily limited by the RF measurement circuitry (~1 Hz). The observed linewidths are predominantly influenced by external noise sources, resulting a Gaussian broadening on the otherwise ideal Lorentzian profile of the sidebands. To account for both effects, we employ a Voigt profile [64]—a convolution of Lorentzian and Gaussian spectra—to fit one of the sidebands, as shown in the inset of Fig. 2(c) [Supplemental Materials S-10]. The fitting yields an effective Voigt linewidth of approximately ~ 0.996 $Hz$, reflecting the combined Gaussian and Lorentzian contributions. This linewidth we obtain is consistent with previously reported values for all-electrical microwave frequency combs [34,38].

We examine the role of the resonant cavity in comb generation. Fig. 2(d) shows $S_{11}$ near the resonance. The insets labelled **I**, **II**, and **III** displays frequency combs generated with a $10\ kHz$ and $0.125\ V_{pp}$ a sine-wave modulation while carrier tone is tuned to the frequencies marked by filled circles matching the labels. Points **I** through **III** correspond to combs generated when the carrier is on-resonance, detuned by $10.47\ MHz$, and off-resonance, respectively. On-resonance (Point **I**) shows significantly higher sidebands and amplitudes compared to off-resonance (Points **II** and **III**). This enhancement results from field confinement and phase modulation due to the cavity Q-factor. Comb generation, though weak, even when detuned (Point **III**) confirms the system's inherent non-linearity. However, a resonant cavity is crucial for comb-generation with extended span and higher amplitudes.

## C. Tunability of the FSR and comb spectrum

A periodic pulse train in the time-domain translates into a discrete structure within the frequency domain, characterized by a *Sinc* function envelope. Fig. 3(a) illustrates this relationship by generating frequency combs using a pulse train applied between the centre conductor and ground plane, for different pulse widths; $\tau = 30\ \mu s$ (top), $15\ \mu s$ (middle) and $5\ \mu s$ (bottom). The schematic representation of pulse trains are shown on the left, with their full frequency spectra in the central panel. A magnified view of the green shaded region in the middle panel is displayed on the right. the frequency combs generated for pulse-widths $30\ \mu s$, $15\ \mu s$, and $5\ \mu s$ exhibit a frequency span between the primary minima, $f_{bw} \approx 66\ kHz$, $133\ kHz$, and $400\ kHz$ respectively. Supplemental Materials S-11 shows a plot of $f_{bw}$ versus $\tau$ fit to the equation $f_{bw} = {}^{2}/_{\tau}$, confirming the reciprocal relationship between the time and frequency domains. These results also highlights the tunability of the frequency comb structure via pulse shaping.

In Fig. 3(b) we explore the amplitude dependence of the frequency-comb using a $10\ kHz$, $15\ \mu s$ modulation pulse with three different peak-to-peak amplitudes $0.0625\ V_{pp}$ (top), $0.250\ V_{pp}$ (middle), and $0.625\ V_{pp}$ (bottom). We find that both number and intensity of the comb-lines increase with the amplitude of the modulation signal; the number of comb-lines increases from 120 to 400 when the modulation amplitude is increased from $0.0625\ V_{pp}$ to $0.625\ V_{pp}$. For the same increment in the modulation amplitude, the SNR for the first side-band also shows a considerable increase, $\sim 12\ dB$ to $\sim 25\ dB$. Higher pulse heights results in stronger phase modulation and subsequently a more robust frequency comb generation.

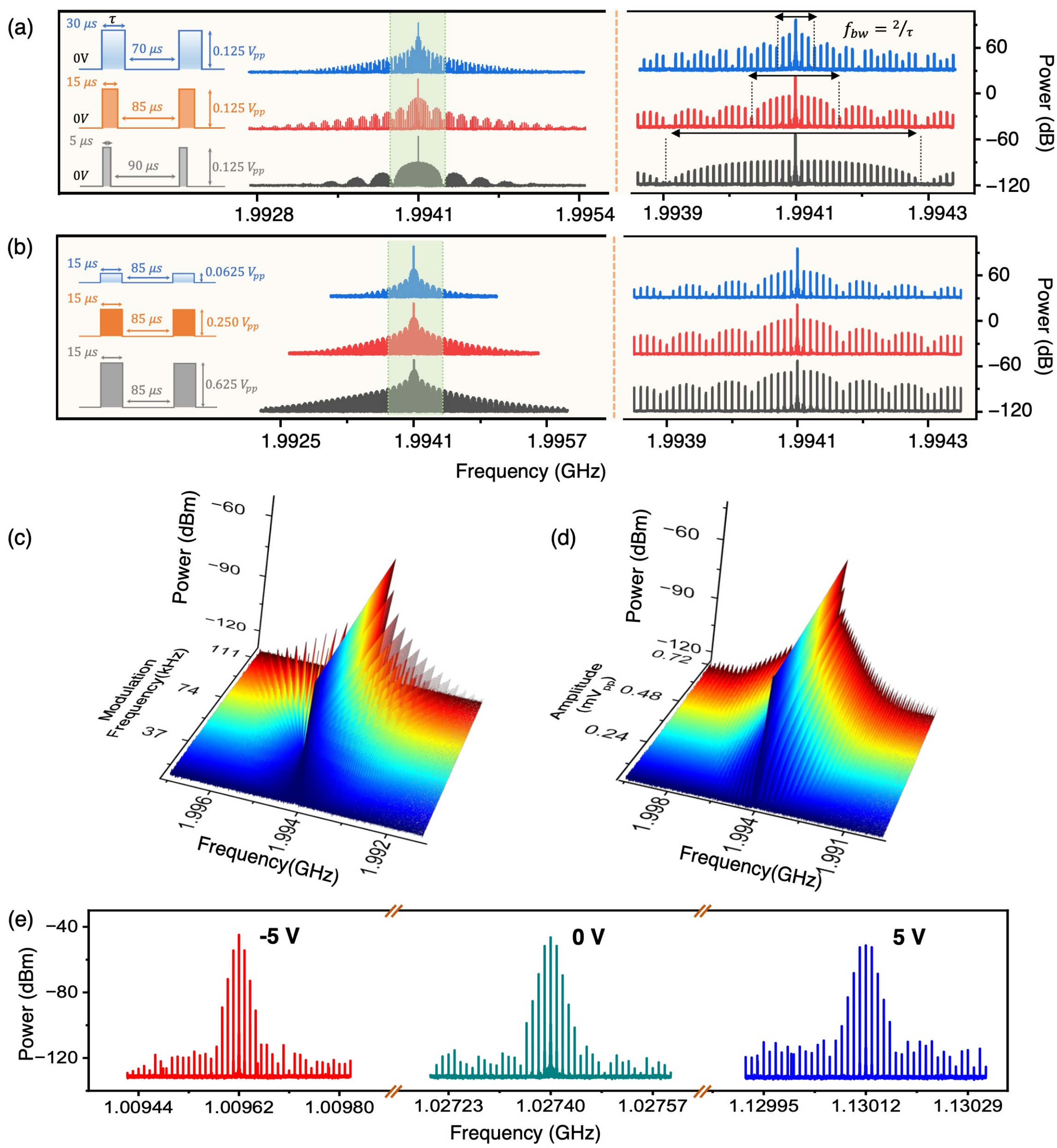

**Fig. 3. Tunability of frequency combs with modulation amplitude and frequency. (a)**, Frequency comb generation using a $0.125\ V_{pp}$ periodic pulse train modulation for three different pulse widths illustrated on the left; $30\ \mu s$ (top), $15\ \mu s$ (middle), and $5\ \mu s$ (bottom). The magnified view of the region enclosed inside the shaded box is shown on the right. **(b)**, Frequency comb generation using three different modulation amplitudes, illustrated on the left — 0.0625 (top), 0.250 (middle), and $0.625\ V_{pp}$ (bottom) with a constant pulse width of 15 µs. The right panel shows a magnified view of the highlighted spectral region. **(c)**, and **(d)** represents the combs generated varying modulation frequency and modulation amplitude: in **c**, a square-wave signal of fixed amplitude ($0.125\ V_{pp}$) is applied with modulation frequencies ranging from $1\ kHz$ to $100\ kHz$. **(d)**, Modulation frequency fixed at $100\ kHz$, while the amplitude is varied from $0.0625\ V_{pp}$ to $0.625\ V_{pp}$. **(e)**, Comb obtained with the application of different DC voltages across the centre conductor and ground-plane for -5 V (left), 0 V (centre) and +5 V (right), demonstrating the tuneability of the spectrum. The data correspond to a 1 GHz resonator driven by a 10 kHz sine-wave modulation.

Fig. 3(a) and 3(b) indicate that for a given FSR, varying the modulation amplitude and pulse widths can control the intensity and number of comb-lines. Figures 3(c) and 3(d) examine the tunability of the comb-lines against the modulation frequency and amplitude, respectively. In Fig. 3(c) the frequency of the square-wave modulation is varied from $1\ kHz$ to $100\ kHz$ while the amplitude is held constant at $0.125\ V_{pp}$. The combs generated for frequencies from $100\ kHz$ to $1\ MHz$ are shown in Supplemental Materials S-12. The comb lines are densely packet at $1\ kHz$, however, increasing the modulation frequency increases the spacing between the comb-lines. The FSR precisely aligns with the applied modulation frequency, confirming the tunability of the comb-lines. In Fig. 3(d), the modulation frequency is maintained at $100\ kHz$, while the amplitude is varied from $0.0625\ V_{pp}$ to $0.625\ V_{pp}$. At lower modulation amplitudes, only a few sidebands with lower amplitudes appear, while for larger amplitudes the number of visible sidebands increase notably, with over 50 sidebands on either side of the carrier for the highest modulation amplitude. Additionally, the power enclosed in the sidebands also rises with the modulation amplitude.

An important aspect of our frequency comb lies in the tunability of the carrier frequency around which the comb is generated; the resonant frequency of the cavity (carrier frequency) can be tuned by the application of a DC voltage, as described in Fig. 1(d). We demonstrate this property using a CPW cavity with a resonant frequency ~ 1. 027 GHz, at zero bias, as shown in Fig. 3(e). When a $-5\ V$ bias is applied, the resonance shifts to a lower frequency(1.0096GHz), conversely a $+5\ V$ bias shifts the entire spectrum toward higher frequencies (1.130 GHz), with an overall shift of about $100\ MHz$ from the zero-bias resonance. These results demonstrate that not only the FSR but also the entire comb spectrum can be tuned—an aspect that is notably absent in other conventional methods of comb generation.

**D. Correlation between the time and the frequency domain responses**

In Fig. 4(a), we demonstrate the frequency comb in the time-domain using another device designed to operate at a resonant frequency of $\sim 370\ MHz$, with a $1\ kHz$, $0.334\ V_{pp}$ sinusoidal modulation (superimposed in red) is applied between the centre conductor and the ground plane. Fig. 4(b) shows the corresponding frequency domain spectra measured with a spectrum analyser. A magnified view of the comb-lines generated is shown in the inset to Fig. 4(b). We find the generation of comb-lines with a precise FSR, corresponding to the modulation frequency, extended over a large frequency span of $\sim 300\ kHz$. It is clear from Fig.

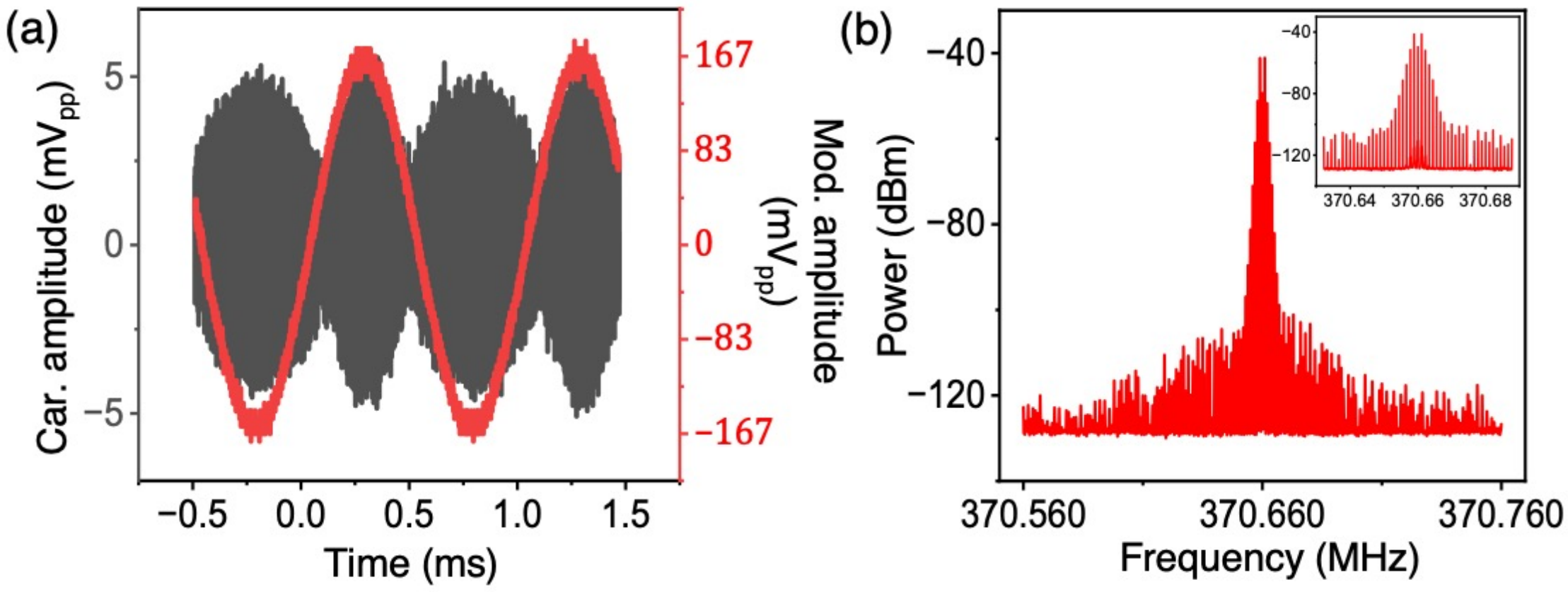

**Fig. 4. Time and frequency domain responses. (a),** Time-domain response of the device under a $0.334\ V_{pp}$ , $1\ kHz$ sinusoidal modulation. The red trace represents the modulation signal, and the black trace is the recorded response from the resonator. **(b),** Corresponding frequency-domain response (FFT) of the signal shown in **(a)**. The inset provides a magnified view of the central frequency region, highlighting the generated comb structure.

4(a) that in addition to the phase modulation, amplitude modulation is also present due to periodic changes in the cavity's electrical length, modulating its reflectance accordingly. We observe an asymmetry between the positive and the negative half-cycles of the modulation envelope in Fig. 4(a). Although the device is lithographically symmetric, the electrical dimensions for the positive and negative half cycle may not symmetric due to the presence of defects and the possible formation of the *dead layer* [57–59], causing memory effects and remnant polarization. The asymmetry observed in Fig. 1 (d) in the frequency response against the electric field arises from the same underlying mechanism.

## III. Conclusions

In conclusion, this study successfully demonstrates the creation of a cryogenic microwave frequency comb by leveraging the effective second-order non-linearity of STO in its quantum paraelectric phase. The comb is generated within a high-Q CPW cavity fabricated on STO, driven by a dual-tone excitation scheme. The large *Q*-factor of the cavity plays a pivotal role in achieving a robust comb generation, as it traps and subjects the microwave signal to phase modulation over multiple cycles. The design offers complete tunability of the resonant frequency, modulation frequency, and FSR. Additionally, the exceptionally high dielectric constant of STO in its quantum paraelectric state reduces the device dimensions by a few orders compared to those built on standard substrates, making it ideal for on chip integration with other circuit quantum devices. In contrast to optical frequency combs [2], which generally offer a broader spectral span, the bandwidth of the comb introduced here is determined by the cavity resonance linewidth—this, in turn, is governed by the overall cavity losses. In our design, both the cavity bandwidth and the operating frequency can be varied by adjusting the cavity dimensions. The central frequency can also be varied by a static electric field. Such microwave frequency combs with a narrow bandwidth are valuable for applications involving frequency mixing, phase modulation, phase shifting, and arbitrary waveform synthesis [65–67], which are central to control and readout of superconducting and semiconducting qubits. Within this framework, a cryogenic microwave frequency comb-based architecture, such as the one described in this manuscript, could be integrated directly with quantum hardware. Such integration has the potential to replace numerous RF sources, thereby significantly reducing wiring complexity, thermal load, and associated noise.

**Acknowledgements:** MT acknowledge funding support from the National Quantum Mission, an initiative of the Department of Science and Technology, Government of India, and PM acknowledges UGC for fellowship.

# Supplemental Materials

## Cryogenic microwave frequency combs based on quantum paraelectric superconducting resonators

Prasad Muragesh[1 #], Harikrishnan Sundaresan [1 #], and Madhu Thalakulam [1 *]

[1] School of Physics, Indian Institute of Science Education & Research Thiruvananthapuram, Thiruvananthapuram Kerala 695551 India.

[#] equally contributed to the work
[*] madhu@iisertvm.ac.in

## S-1. A short survey on the reported microwave frequency combs.

| S/N | Title of publication/journal/volume, year. | DOI | Centre frequency | Spectrum span | FSR | Operating temperature | Source | Tuneable |
|---|---|---|---|---|---|---|---|---|
| 1 | Optomechanical dissipative solitons. *Nature* **volume 600**, pages75–80 (2021) | https://doi.org/10.1038/s41586-021-04317-1 | ~ 17.6 MHz | 9 MHz | 500kHz | RT | | No |
| 2 | Magnetics frequency comb in the magnomechanical resonator. Phys. Rev. Lett. 131, 243601 (2023) | DOI: https://doi.org/10.1103/PhysRevLett.131.243601 | ~ 4.725GHz | 250MHz | 120 to 125MHz | RT | Microwave signal generator | No |
| 3 | Bifurcation generated mechanical frequency comb. Phys. Rev. Lett. 121, 244302 (2018) | DOI: https://doi.org/10.1103/PhysRevLett.121.244302 | | 10Hz | 0.5Hz | | Mechanical resonator driven by a single actuation signal | No |
| 4 | Generation of Soliton Frequency Combs in NEMS. Nano Lett. 2024, 24, 10834–10841 | https://doi.org/10.1021/acs.nanolett.4c02249 | | 8MHz (1st soliton), 30MHz(multi-soliton) | 10kHz | RT | CW Laser | No |
| 5 | NEMS generated electro-mechanical frequency combs. Microsystems & Nanoengineering volume 11, Article number: 8 (2025) | https://doi.org/10.1038/s41378-024-00860-9 | | 30MHz | 101.1kHz and 202.2kHz | RT | CW Laser (5mW) | No |
| 6 | Optomechanical frequency comb based on multiple nonlinear dynamics Phys. Rev. Lett. 132, 163603 (2024) | DOI: https://doi.org/10.1103/PhysRevLett.132.163603 | | 12GHz | 77.19MHz and 90.17MHz | 50 mK | Laser | No |
| 7 | Frequency comb in a parametrically modulated micro-resonator. Acta Mech. Sin. 38, 521596 (2022) | https://doi.org/10.1007/s10409-022-21596-x | ~ 2.59MHz | 120Hz | 12 and 18Hz | RT | | No |
| 8 | Integrated and DC powered superconducting microcombs. Nature Communications volume 15, Article number: 4009 (2024) | https://doi.org/10.1038/s41467-024-48224-1 | | 14.97 GHz | 750 MHz | low temperature (mK) | Microwave lasing | No |
| 9 | Comb generation in superconducting resonator. Phys. Rev. Lett. 113, 187002 (2014) | DOI: https://doi.org/10.1103/PhysRevLett.113.187002 | | 8GHz | 60kHz to 1.2MHz | 3 K | Microwave signal generator | Yes (Only FSR can be tuneable) |
| 10 | Tuneable electromechanical comb generation. Appl. Phys. Lett. 100, 113109 (2012) | https://doi.org/10.1063/1.3694041 | ~ 2.8 MHz | 25Hz | 5 Hz to 0.005Hz | RT | electromechanical | Yes (Only FSR can be tuneable) |
| 11 | Controllable microwave frequency comb generation in a tunable superconducting coplanar-waveguide resonator Chinese Physics B, Volume 30, Number 4 (2021) | DOI: 10.1088/1674-1056/abc2bb | 5.757GHz | 100MHz | 5Hz, 5kHz, and 5MHz | 20 mK | | Yes (Only FSR can be tuneable) |
| **12** | **This work** | | **Tuneable** | **Tuneable** | **Tuneable** | **few mK to <1.2K** | **Microwave signal generator** | **Yes (Both centre frequency and the FSR tuneable)** |

### S-2: Measurement setup

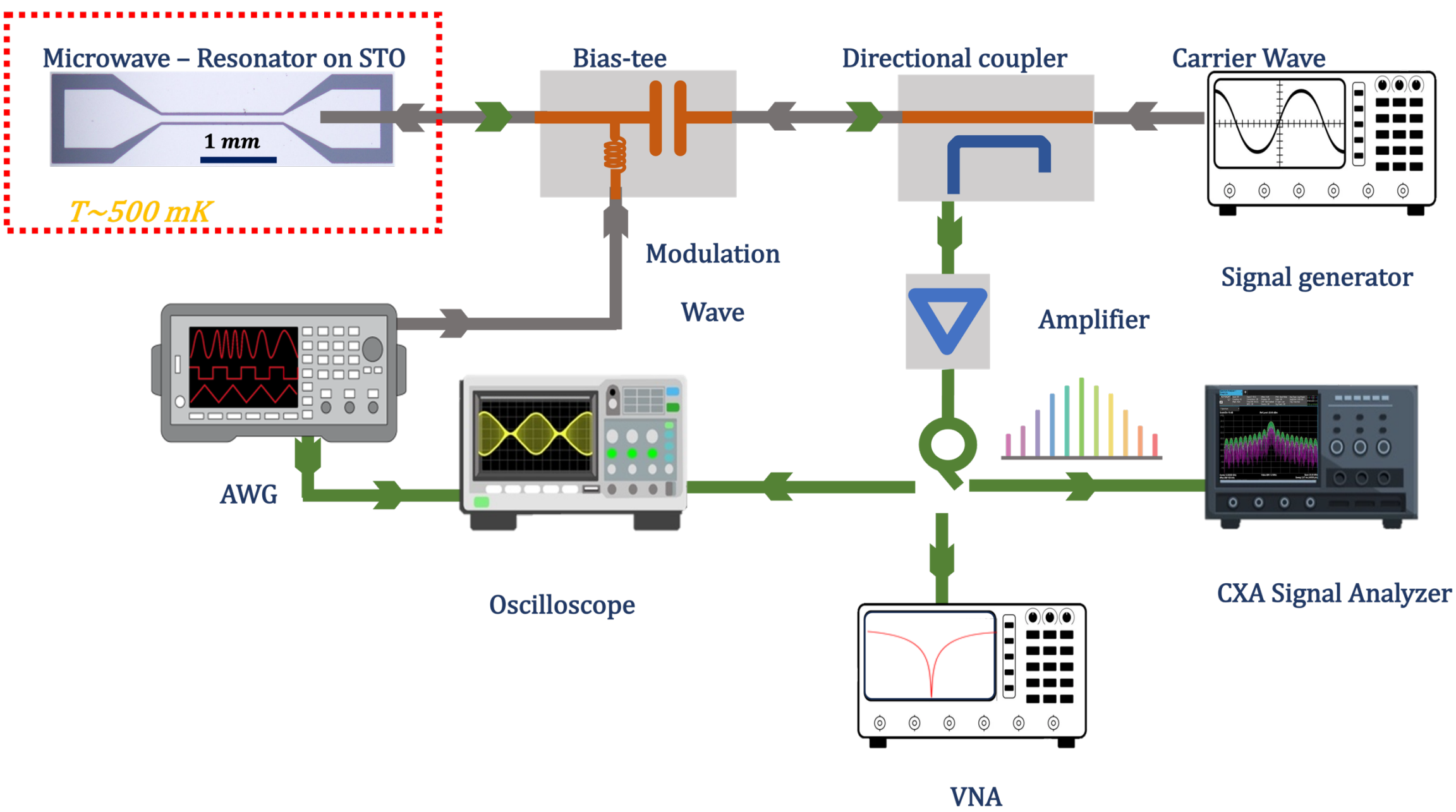

All the electrical circuitries are installed at room temperature, except for the resonator, which is mounted on the 500 mK stage of a cryogen-free dilution refrigerator. The device is wire-bonded to 50 Ω impedance connecting pads terminated by SMA connectors. High-frequency coaxial cables were used to interface the device with the room-temperature electronics, ensuring minimal signal degradation. All the measurements are done in the RF-reflectivity mode with the help of a directional coupler. A bias-tee is used to add the modulation and carrier wave signals. The resonant frequencies identified using a Vector Network Analyzer and a DC voltage source. For realizing the frequency-comb, the carrier tone is supplied from an RF signal generator while the modulation tone is generated using an arbitrary waveform generator (AWG). The reflected signal from the device is amplified using a low-noise amplifier with a 13 $dB$ gain and is then routed either to a spectrum analyser for frequency-domain analysis or to an oscilloscope for time-domain measurements.

**S-3: Device fabrication**

The $\lambda/2$ superconducting CPW resonators with a length $1.5\,mm$, centre conductor width $30\,\mu m$ and the ground-plane separation $10\,\mu m$ is designed to have a characteristics impedance $1\,\Omega$ and fundamental resonant mode $\sim 1\,GHz$, at a temperature below 4 K, assuming a dielectric constant of ~ $20{,}000\ to\ 24{,}000$. The resonator is fabricated on a commercially available single crystal (001), $500\,\mu m$ thick STO substrate using standard photolithography.

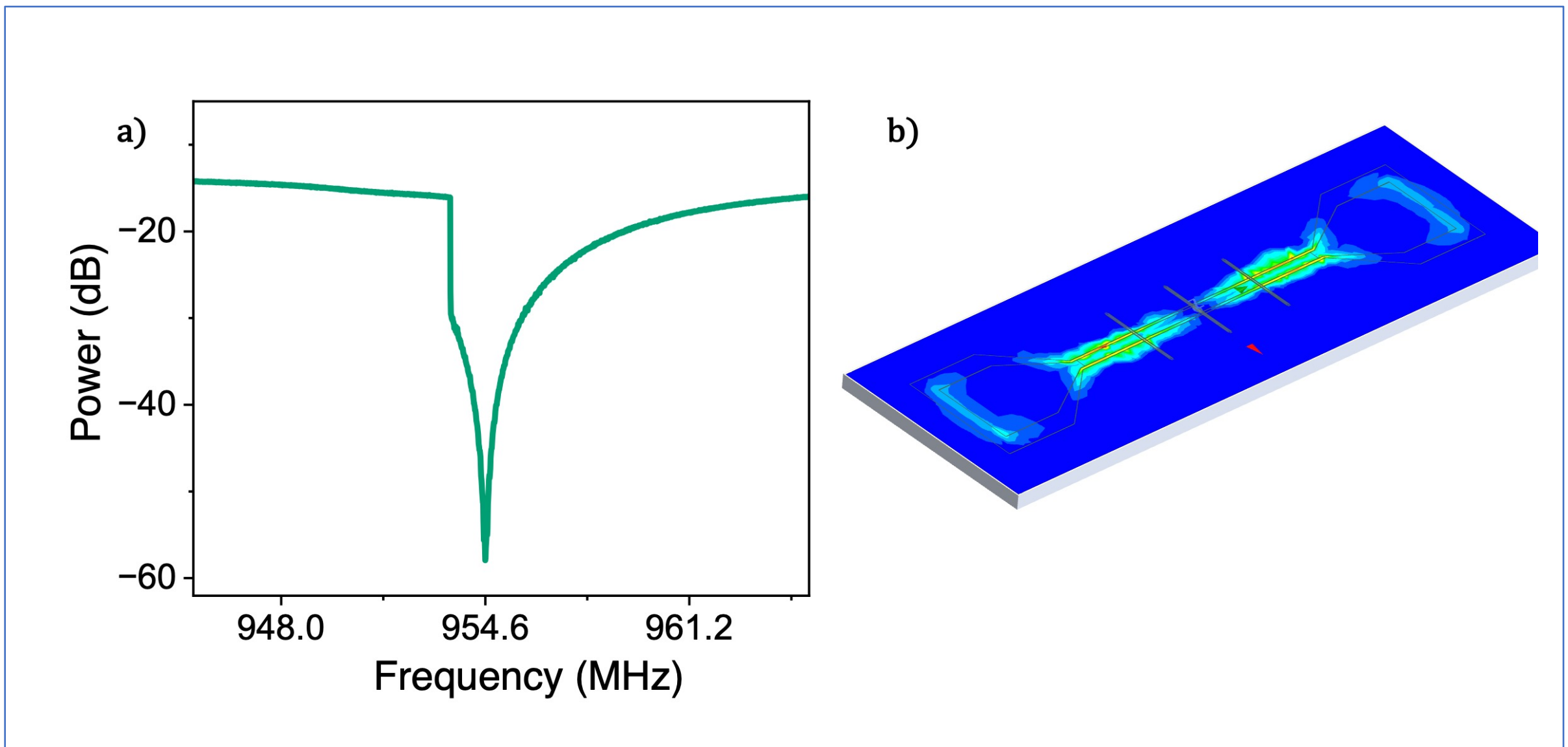

**S-4.** (a) The measured $S_{11}$ of fundamental resonance mode around 955 MHz. (b) Simulated electric field profile of the fundamental mode obtained using the eigen-mode analysis ( Ansys HFSS).

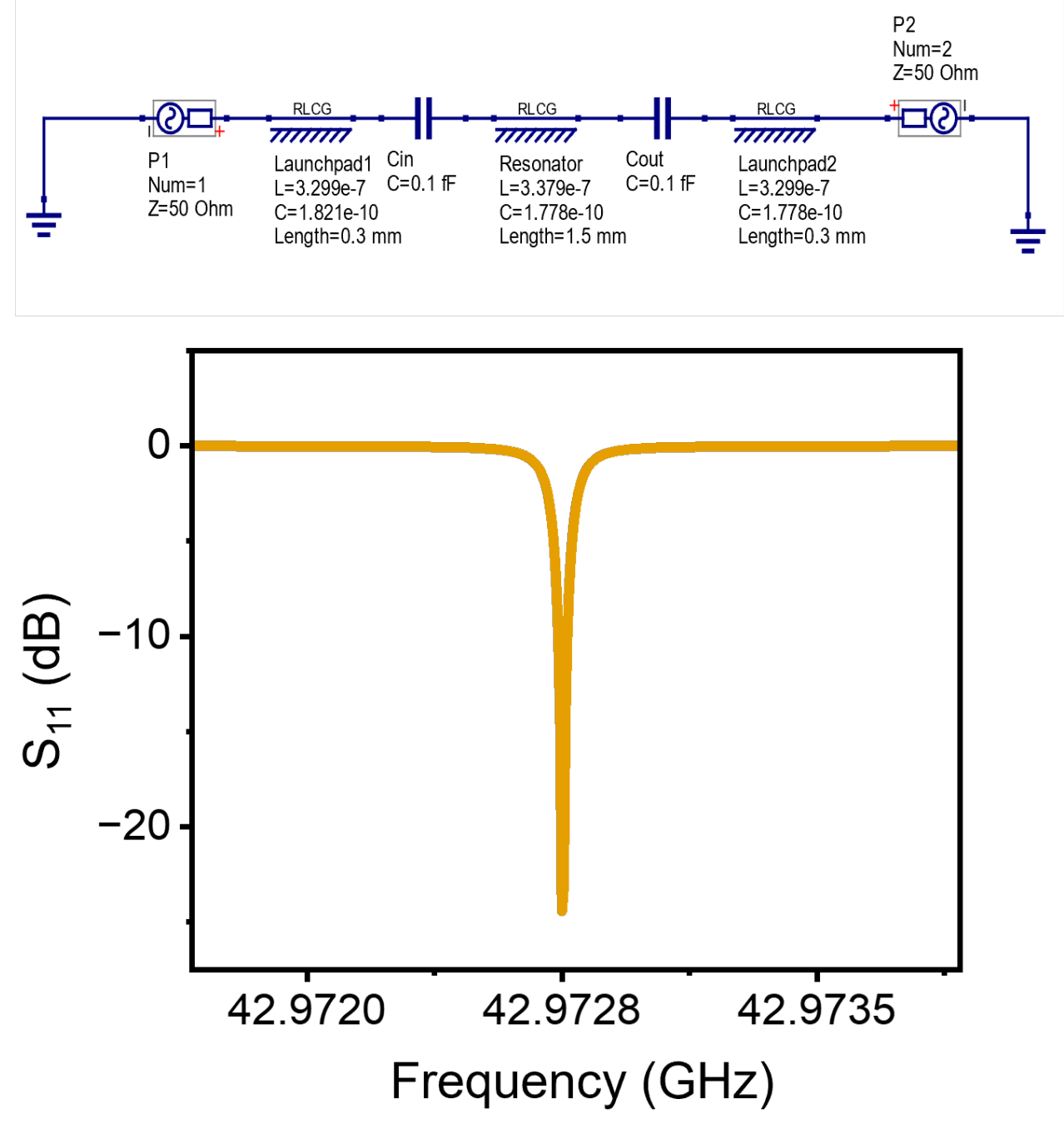

**S-5. Calculation of dielectric constant of STO**: schematic of Simulated circuit and $S_{11}$ of fundamental resonance mode on sapphire substrate using the QUCS. To extract the effective dielectric constant of STO, we employed a comparative approach using theoretical calculations and electromagnetic simulations. We simulated the identical resonator geometry on a sapphire substrate, which has a well-known dielectric constant of $\varepsilon_{eff}^{Al_2O_3} \sim 5.15$ to 6.25, and obtained a fundamental resonance frequency of approximately $f_1^{Al_2O_3} = 43$ GHz. The figure below shows the simulated S11 response obtained using QUCS for an identical resonator on a sapphire substrate. Since the inductive component of the resonator remains the same across both configurations, the difference in resonance frequencies can be attributed solely to the variation in the effective dielectric permittivity. By comparing the simulated frequency $f_1^{Al_2O_3}$(on sapphire) with the experimentally measured frequency $f_1^{STO}$ on STO, we extracted the effective dielectric constant of STO using the relation.

$$\varepsilon_{eff}^{STO} = \varepsilon_{eff}^{Al_2O_3} \left(f_1^{Al_2O_3}/f_1^{STO}\right)^2 \qquad \text{(i)}$$

## S-6. The relative phase shift (δ)

The equation (3) describes the **phase shift** that occurs when a wave travels through a medium where the refractive index modulated.

The phase (∅) of a wave traveling a distance $L$ through a medium is defined by the propagation constant $\gamma$ and the length.

$$\emptyset = \gamma \cdot L$$

In a medium, the propagation constant $\beta$ is related to the free-space wavelength ($\lambda$) and the **refractive index ($n$)** of that medium.

$$\gamma = \frac{2 \cdot \pi}{\lambda} \cdot n$$

The dielectric constant $\sqrt{\varepsilon_{eff}}$ in our system is analogues to the refractive index ($n$), replace $n$ with $\sqrt{\varepsilon_{eff}}$ from the above equation. This leads to

$$\gamma = \frac{2 \cdot \pi}{\lambda} \cdot \sqrt{\varepsilon_{eff}}$$

Phase accumulated over the length $L$ is given by

$$\emptyset = \frac{2 \cdot \pi}{\lambda} \cdot \sqrt{\varepsilon_{eff}} \cdot L$$

With the application of bias across the cavity the dielectric constant changes leads to the change in the permittivity.

The relative phase shift against variation in the dielectric constant

$$\delta = \frac{2 \cdot \pi}{\lambda} \cdot L \left(\sqrt{\varepsilon_{eff,2}} - \sqrt{\varepsilon_{eff,1}}\right)$$

So,

$$\delta = \frac{2\pi}{\lambda} \Delta \left(\sqrt{\epsilon_{eff}}\right) \cdot L$$

Where

$$\Delta \left(\sqrt{\varepsilon_{eff}}\right) = \sqrt{\varepsilon_{eff,2}} - \sqrt{\varepsilon_{eff,1}}$$

| Sl no | Carrier wave phase noise without DUT | | Carrier wave phase noise with DUT | | phase noise of 1st side band | |
|---|---|---|---|---|---|---|
| | Frequency | Phase noise | Frequency | Phase noise | Frequency | Phase noise |
| **1** | 1 Hz | -63.82 dBc/Hz | 1 Hz | -64 dBc/Hz | 1 Hz | 0 dBc/Hz |
| **2** | 10 Hz | -63.92 dBc/Hz | 10 Hz | -67.69 dBc/Hz | 10 Hz | -71.07 dBc/Hz |
| **3** | 100 Hz | -87.03 dBc/Hz | 100 Hz | -97.88 dBc/Hz | 100 Hz | -84.89 dBc/Hz |
| **4** | 1 kHz | -108.02 dBc/Hz | 1 kHz | -110.39 dBc/Hz | 1 kHz | -85.7 dBc/Hz |
| **5** | 10 kHz | -110.79 dBc/Hz | 10 kHz | -112.11 dBc/Hz | 10 kHz | -100 dBc/Hz |
| **6** | 100 kHz | -108.4 dBc/Hz | 100 kHz | -111.84 dBc/Hz | 100 kHz | |
| **7** | 1 MHz | -117.7 dBc/Hz | 1MHz | -122.71 dBc/Hz | 1MHz | |
| **8** | 10MHz | -114.4 dBc/Hz | 10MHz | -120 dBc/Hz | 10 MHz | |

**S-7. Spectral Purity:** The table presents the spectral purity (phase noise) values extracted from Fig. 2(b) of the main article. It provides the phase noise data for the carrier wave without the DUT (device under test), the carrier wave with the DUT, and the first sideband, listed for each decade offset frequency.

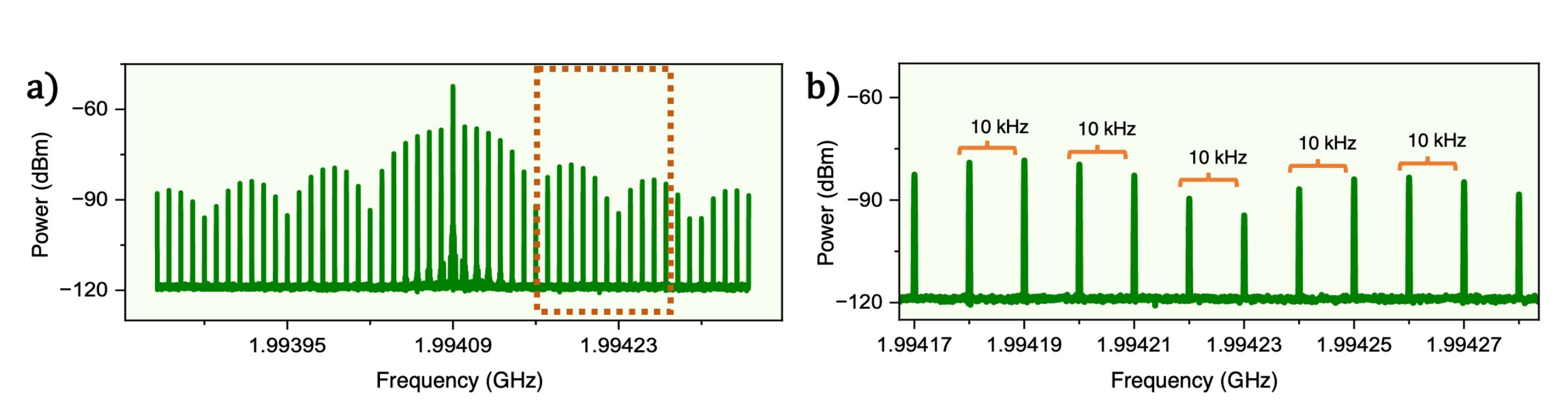

**S-8. Equidistant Spacing of Comb Lines:** The data correspond to a 10 kHz pulse modulation signal with a pulse width of 15 µs and an applied voltage of 0.625 $V_{pp}$ between the centre conductor and ground plane. The zoomed image on the right highlights the region marked by the orange dotted box in the left figure and clearly shows the equidistant 10 kHz spacing between the sidebands.

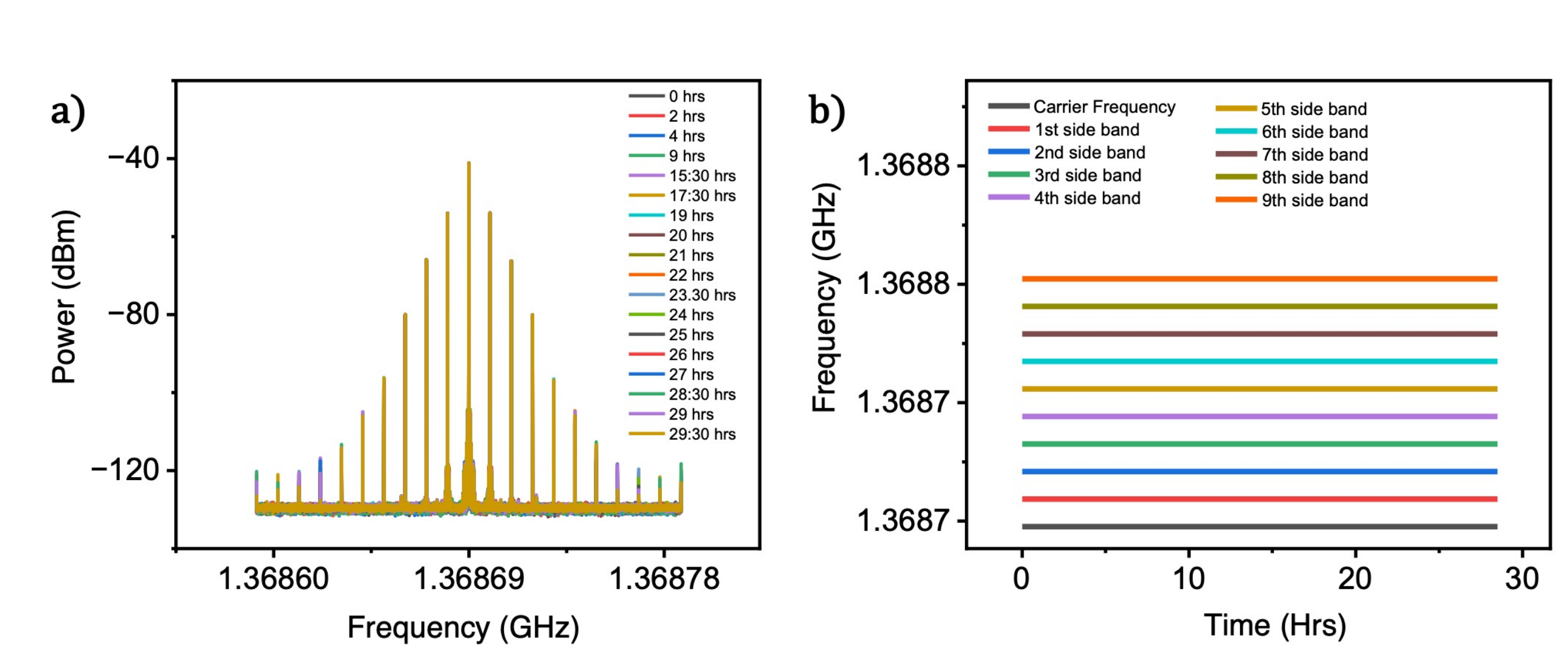

**S-9. Drifting of Comb Lines Over Time**: Figure (a) shows the comb spectrum generated from a 1.36 GHz resonator using a sine wave as the modulation signal. The comb data were recorded over a period of ~30 hours to monitor possible frequency drifts in the spectral lines. Figure (b) presents the frequency vs. time plot extracted from Fig. (a), demonstrating that the comb lines remain highly stable with no observable drifting over time.

**S-10 Voigt profile**

Voigt profile is the convolution of a Gaussian profile, $G(x;\sigma)$ and a Lorentzian profile, $L(x;\gamma)$:

$$V(x;\sigma,\gamma) = \int_{-\infty}^{\infty} G(x';\sigma)L(x-x';\gamma)dx' \quad (1)$$

Where

$$G(x;\sigma) = \frac{1}{\sigma(\sqrt{2\pi})}\exp\left(-\frac{x^2}{2\sigma^2}\right) \text{ and } L(x;\gamma) = \frac{\gamma/\pi}{x^2+\gamma^2}$$

The full width half maxima of (FWHM) of the Voigt profile can be found from the width of the associated Gaussian and Lorentzian widths.

The FWHM of Gaussian profile is

$$f_G = 2\sigma\sqrt{2\ln(2)}$$

And FWHM of Lorentzian profile is

$$f_L = 2\gamma$$

An approximate relation between the widths of the Voigt, Gaussian, and Lorentzian profile is

$$f_V = f_L/2 + \sqrt{f_L^2/4 + f_G^2} \quad (2)$$

The extracted values of $f_L$ and $f_G$ Voigt fit from the figure 2(c) inset from the main article are.

$$f_L = 2.4546 \times 10^{-17} \pm 3.84 \times 10^{-5} \text{ and } f_G = 0.99249 \pm 3.3086 \times 10^{-4}$$

The calculated $f_V$ from equation (2) with the above values inserted

$$\boldsymbol{f_V = 0.9962}$$

The Extracted line width of side band shown in figure 2(c) inset from the main article is $\sim \mathbf{0.9962}$

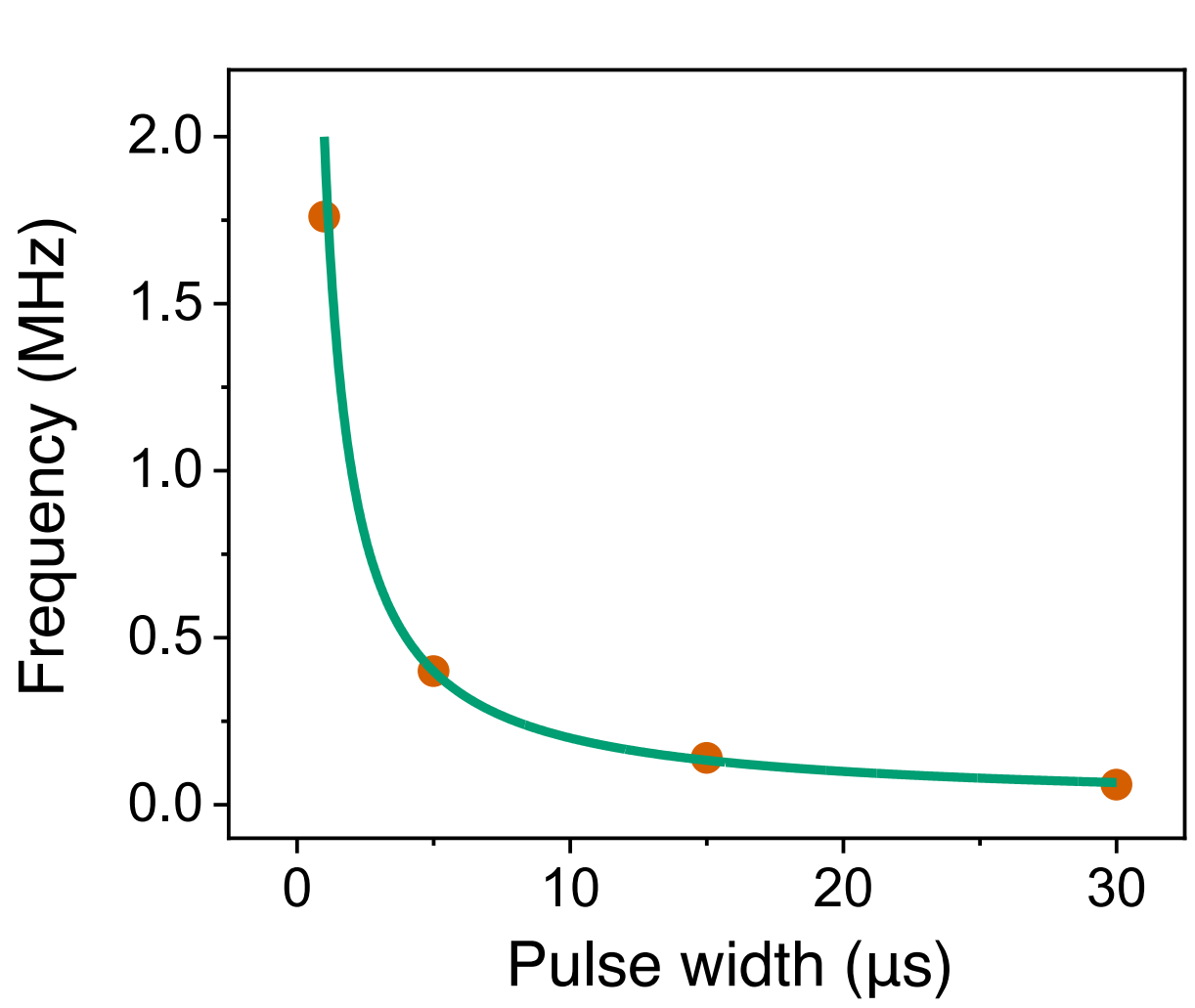

**S-11. Pulse width $\tau$ vs $f_{bw}$:** The filled circles represent experimental values while the green line is a fit to the equation $f_{bw} = {}^{2}/_{\tau}$

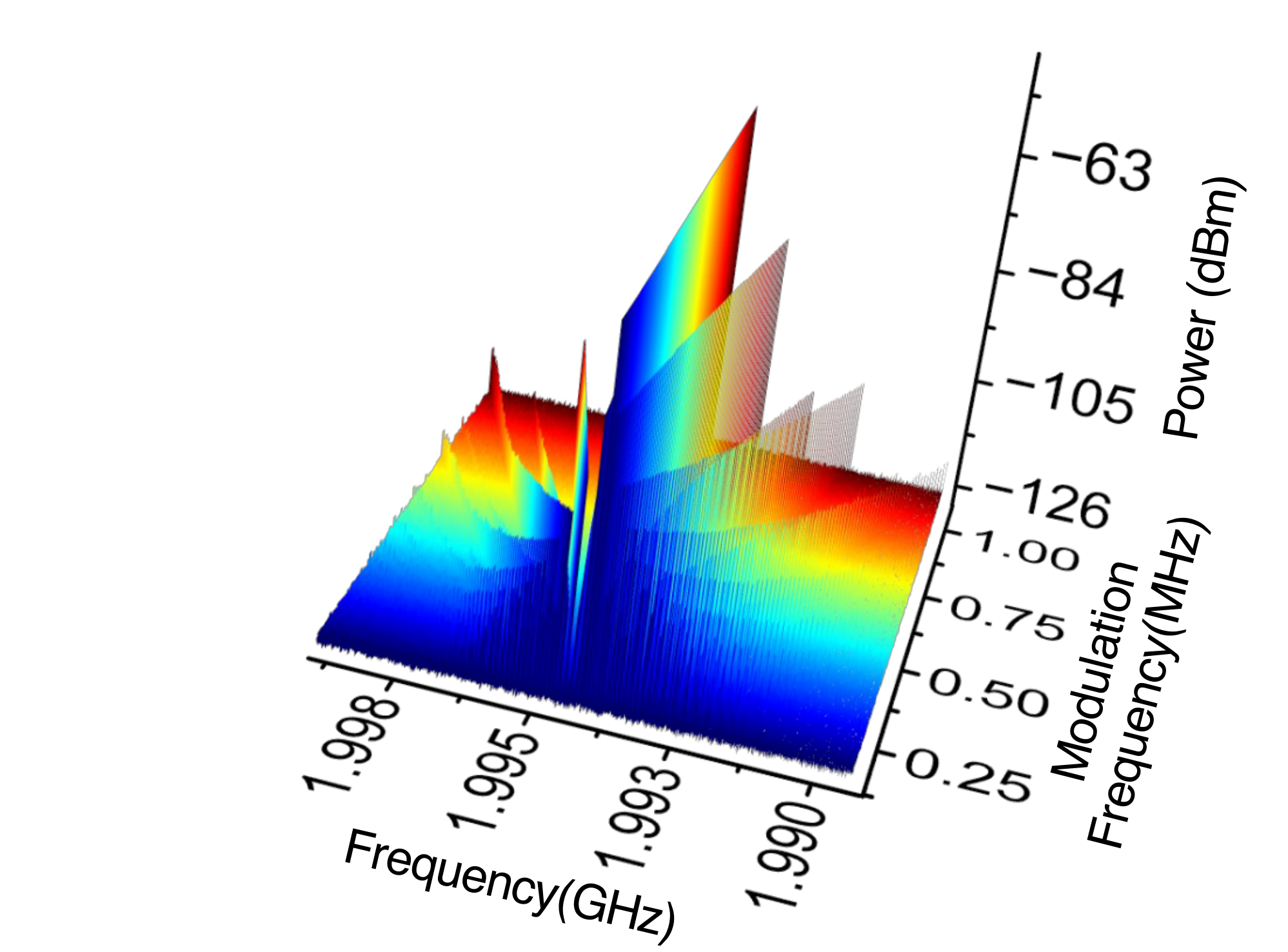

**S-12. 2D plot of frequency dependent study from 100kHz to 1MHz**: Plot of the generated frequency comb vs modulation frequency, demonstrating the tunability of the comb-lines using a 125 mVpp square-wave modulation in the 100 kHz up to 1 MHz range.